# Hiding Sound in Image by k-LSB Mutation

Ankur Gupta
Dept. of Computer Science
Birla Institute of Technology & Science
Pilani, India
f2009087@pilani.bits-pilani.ac.in

Ankit Chaudhary, Member IEEE
IA, USA
dr.ankit@ieee.org

*Abstract*— In this paper a novel approach to hide sound files in a digital image is proposed and implemented such that it becomes difficult to conclude about the existence of the hidden data inside the image. In this approach, we utilize the rightmost k-LSB of pixels in an image to embed MP3 sound bits into a pixel. The pixels are so chosen that the distortion in image would be minimized due to embedding. This requires comparing all the possible permutations of pixel values, which may would lead to exponential time computation. To speed up this, Cuckoo Search (CS) could be used to find the most optimal solution. The advantage of using proposed CS is that it is easy to implement and is very effective at converging in relatively less iterations/generations.

I. INTRODUCTION

As internet accessibility is to everyone and anywhere, there has been an enormous spike in eavesdropper and information stealing. To safely transmit data through the internet, some mechanisms must be provided to guard important data against illegal interception. Many cipher algorithms have been proposed for this where user need to encrypt data before transmission and decrypt the same after reception [1-3] and data are protected from illicit access. Although data encryption is a de facto standard to secure data but there is still some scope for improvement. The appearance of cipher texts would give unauthorized user an impulse to recover them. Moreover, the unauthorized users might even simply destroy the cipher text out of range when they have trouble recovering them so that the legal receivers cannot get the data in time or will get a wrong one. Therefore hiding the mere presence of data has been an active topic of research [4].

Data hiding is the technique of embedding data into audio or visual media, called host signal such as music, images, and videos. It can be utilized in a wide variety of applications such as ownership identification, tamper proofing, caption, and secret data transmission. Unlike the goal of data encryption that prevents the data from illicit access and modification, the aim of data/information hiding is to make the data inaudible or invisible to the grabbers. Watermarking and steganography are two major branches of data/information hiding technology. Both have their specific pros and cons. In Watermarking, a distinguishable symbol e.g., a signature is embedded into signals to authorize the ownership of the signals. Usually a small sized symbol, ranging from one bit to thousands of bits is used.

On the other hand, Steganography is the art of hidden writing. The word Steganography comes from the Greek words *steganos* and *graphia*, which together means "hiding writing" [5]. Covert communication aimed at hiding a message from a third party, is the main purpose of steganography. This differs from cryptography, which is intended to make a message unreadable by a third party but does not hide the existence of the secret communication.

## II. BACKGROUND

In the past two decades, there are numerous least-significant-bit-based (LSB-based) data hiding techniques. Generally they find some pixels based on specialty and suitability of those pixels to act as substitute hosts in the host media and then embed data into the LSB of those pixels [6-9]. Wang et al. [10] proposed to embed secret messages in the moderately significant bit of the cover-image using GA for optimal substitution matrix. Use of local pixel adjustment process (LPAP) was also proposed by him to improve the image quality of the stego-image. He also developed the optimal k-LSB substitution method to solve the problem when k is large. Lil [5] proposed a steganographic method based on JPEG and Particle Swarm Optimization (PSO). This strategy could be used in spatial domain and thus applied to transform domain.

Recently, Bedi [11] proposed the use of PSO to solve the problem of finding best pixels for substitution. PSO is used to find the best pixel locations in a gray scale cover image, where the secret gray scale image pixel data can be embedded. This technique though works well but due to improved solution space, search algorithms as cuckoo search can now give better results. Gerami [12] used PSO coupled with Optical Pixel Adjustment to enhance the image quality. Marghny [13] proposed K-LSB substitution with permutable keys. He used Gene Expressing Programming (GEP) to determine the best permutation of the key for LSB substitution. However none of these works have explored the possibility of hiding a sound file inside an image.

In this work, we propose to use a novel technique using cuckoo search for finding the optimal output. The paper has been organized as follows. The concept of breaking sound files into byte streams has been discussed in Section 3. In Section 4, a technique involving LSB substitution and optimal pixel selection using cuckoo search is developed. Experimental results are given in Section 5, and a brief conclusion is made finally in Section 6.

## III. DECODING AUDIO TO BYTE STREAM

The structure of an MP3 file can be expressed by the following scheme:

ID3v2   Frame1   Frame2   Frame3 …. ID3v1

An mp3 file is divided into small blocks, called frames. Each frame has a constant time length of 0.026 sec. But the size of one frame varies according to bitrate. For example, for a 128kbps signal it is (normally) 417 Bytes and for 192kbps 626 Bytes. The first 4 Bytes of each frame is frame header and the rest is audio data. MP3 structure is shown in Figure 1.

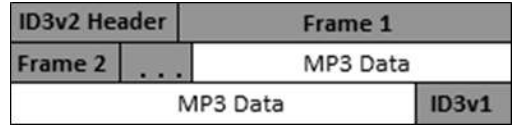

Figure 1 : Structure of an mp3 file.

Frame header consists of information about frame (bitrate, stereo mode) and therefore each frame is an independent entity. Each of them can have its own characteristic. It is used in Variable Bitrate files, where each frame can have different bitrate.

A frame header has the following bit structure:

AAAAAAAA  AAABBCCD  EEEEFFGH  IIJJKLMM

Where A stands for Frame synchronizer, B means MPEG version ID, C means Layer, D stands for CRC Protection, E stands for Bitrate index, F means sampling rate frequency index, G is Padding, H is Private bit, I is Channel, J means Mode extension (only if Joint Stereo is set), K is Copyright, L is Original, and M stands for Emphasis.

Thus MP3 file is decoded into a stream of bytes and the resulting byte stream is attempted to be hidden inside a cover image. While reception, these bytes are encoded back according to byte sequence above and results into a playable audio file with its headers and metadata intact.

## IV. PROPOSED METHOD IN SPATIAL DOMAIN

The Image is divided into n non-overlapping square windows each of size 16 pixels [11]. Each of these square windows can be called a "nest". Since we are substituting a pixel value with its own mutant where its k rightmost LSB are now k bits of message, an egg or a solution is the new mutant pixel. The egg laid by a cuckoo is discovered by the host with a probability $p_a$. we assumed $p_a=0.25$ for the purpose of study. That mean an unfit solution will be discovered with a probability of $p_a$.

The objective needs to take care primarily of two things: low distortion and high robustness of the output image. There are many ways to gauge the distortion in an image, most popular of them being Peak Signal to Noise Ratio (PSNR) and Structural Similarity Index (SSIM). Although PSNR has been the single most popular choice in many research works, in this study we prefer SSIM and PSNR both so that the algorithm is robust to each of these metrics. PSNR becomes more relevant when we are dealing with information loss over an encoding/decoding routine rather than subjecting it to different attacks of varied types [14].

On the other hand, SSIM is more advanced and is primarily developed as an improvement to the PSNR. SSIM is found to be more in line with human perception of pictures than traditional methods such as PSNR and MSE. SSIM measure between two pixel windows, say x and y of same size, say N×M. Formally it is defined as:

$$\text{SSIM}(x, y) = \frac{(2\mu_x\mu_y+c_1)(2\sigma_{xy}+c_2)}{(\mu^2_x+\mu^2_y+c_1)(\sigma^2_x+\sigma^2_y+c_2)} \quad (1)$$

where $\mu_x$, $\mu_y$ are the statistical mean of pixel values in image x and y respectively. $\sigma^2_x$ and $\sigma^2_y$ are the variance of x and y respectively and $\sigma_{xy}$ is the covariance of x and y. c1 and c2 are SSIM constants. SSIM concerns itself with the apparent difference in structural information. The value always remains within [-1, 1], where 1 indicates perfectly matching images.

The PSNR of a given component is the ratio of the maximum mean square difference of component values that could exist between the two images (a measure of the information content in an image) to the actual mean square difference for the two subject images. So the higher the PSNR, the closer the images are. A luminance PSNR of 20 means the mean square difference in the luminance of the pixels is 100 times less than the maximum possible difference, i.e. 0.01. Because many signals have a very wide dynamic range, PSNR is usually expressed in terms of the logarithmic decibel scale. It is expressed as a decibel value.

It is most easily defined via the mean squared error (MSE) which for two M×N monochrome images I and K where one of the images is considered a noisy approximation of the other is defined as:

$$\text{MSE} = \sum_{i=0}^{m-1}\sum_{j=0}^{n-1}[I(i,j)-K(i,j)]^2$$

$$\text{PSNR} = 10.\log\left(\frac{\text{MAX}_I^2}{\text{MSE}}\right)$$

$$= 20.\log(\text{MAX}_I) - 10.\log(\text{MSE}) \qquad (2)$$

Here, $\text{MAX}_i$ is the maximum possible pixel value of the image. When the pixels are represented using 8 bits per sample, this is 255. More generally, when samples are represented using linear PCM with B bits per sample, $\text{MAX}_i$ is $2^B - 1$.

Therefore our fitness function can be formally defined as

$$Z(x,y) = \alpha \times \text{SSIM}(x,y) + (1-\alpha)\text{PSNR}(x,y)$$

Since y is the original image and remains constant during the course of embedding, the above equation can be re-written as

$$Z(x) = \alpha \times \text{SSIM}(x, x_{\text{orig}}) + (1-\alpha)\text{PSNR}(x, x_{\text{orig}})$$

where α is the weight constant. For the purpose of this study $\alpha = 0.5$.

When generating new solutions x (t+1) for, say, a cuckoo I, a Lévy flight is performed:

$$x(i, t+1) = x(i,t) + \alpha \times \text{Lévy}(\lambda)$$

where α > 0 is the step size which should be related to the scales of the problem of interests. In most cases, we can use α=1. The above equation is essentially the stochastic equation for random walk. The product × means entry wise multiplications. An operator that will be used in the novel approach presented is 'mutation' denoted hereby with the symbol '~'. A ~ B[i : j] signifies substituting k-LSB of A with k bits of B, where k is the bit string of B lying between index i and j of B, both inclusive.

The image will go through many changes through the course of this algorithm, with an image from previous generation being modified in the current generation of the presented algorithm. Therefore, following a notation for representing images gains importance. $x_{(t, ck_{\text{space}})}$ represents the modified input image $x_{(i,t)}$ in a state 't'. 'ck_space' is a set of pixel values where the modification has been done, as a result of the algorithm. Also by extension, $x_{(t0,0)}$ represents the input image $x_{(t,i)}$

One round of cuckoo search algorithm consists of selecting nests at random, employing of Lévy flights for substituting the solutions in a nest, followed by abandoning of relatively unfit solutions. Each "egg" is tested while discarding some based on lack of fitness and sending the successful solutions through a second round and so on until an optimal solution emerges. This is carried on until maximum numbers of generation are reached or the stopping criterion is met. Stopping criterion is usually taken to be as convergence of the fitness function.

Our approach treats each 16x16 pixel window in the image as a nest. The nests are randomly picked to be mutated with sound bits. At the end of this algorithm, we will have the mutant cover image with least possible distortion for the given sound to be hidden inside the given image.

## V. EXPERIMENTAL RESULTS

As can be seen from Table 1, the proposed method performs better than the contemporary methods that adopt similar techniques as well as traditional methods that act as benchmarks. Input images used are one of the most popular images in image processing so that results can be easily corroborated.

TABLE I. VALUES OF THE OBJECTIVE FUNCTION Z, OBTAINED AFTER EMBEDDING SOUND BYTES STREAM IN DIFFERENT COVER IMAGES

| Host Image | Simple LSB substitution | Lin [15] (non-PSO method) | Bedi [11] using PSO | Proposed method using cuckoo search |
|---|---|---|---|---|
| Baboon | 44.31 | 43.31 | 45.19 | 46.21 |
| Cameraman | 44.43 | 43.31 | 45.19 | 46.75 |

It is also important to notice that there is no spatial effect in the images after embedding sound data.

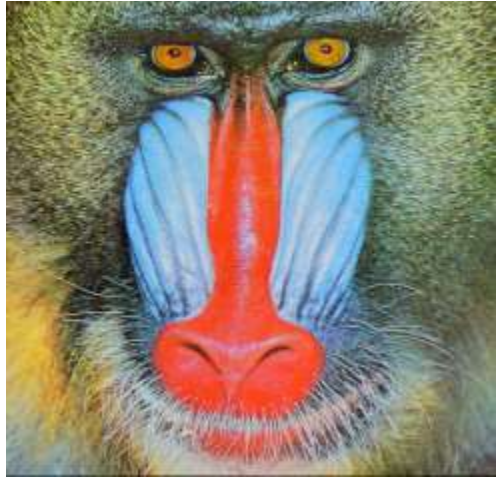

Figure 2. Result Image: Baboon

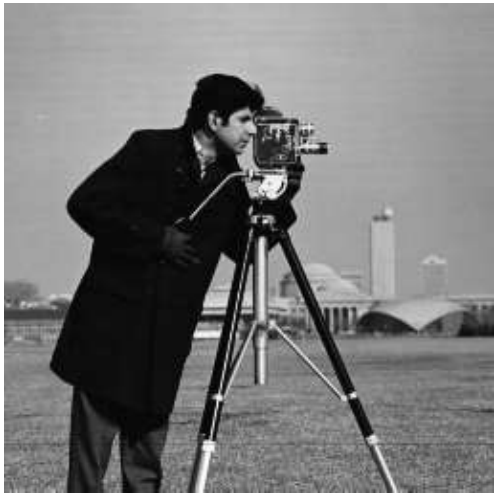

Figure 3. Result Image: Cameraman.

## VI. Conclusion

A novel technique of hiding sound files inside unsuspecting image files has been presented where with minimal distortion in the image quality were acquired. Cuckoo search has been an instrumental tool in this work, used to find the optimal solution set for the given problem. Cuckoo search has reduced the time required to reach at the best solution due to its aping behavior of cuckoo laying behavior therefore using nature's evolutionary techniques to its advantage. Like most of the pixel substitutions algorithms, we have made sure that size of the image does not increase as a result of embedding sound into the image. The proposed method outperforms most state of the art methods wherein it is attempted to hide any binary data in an image, in terms of time efficiency at reaching the optimal/sub-optimal solution and also at introducing minimal distortion to the input signal.